\documentclass[preprint,journal]{vgtc}

\title{ProbeScout: Visual Analytics for Attribute-Guided Image Search}

\preprinttext{Preprint}
\manuscriptnote{}

\author{Yifan Lv, Yiyun Chen, Daojun Ye, Haotian Yang, and Weikai Yang}
\authorfooter{
  \item Yifan Lv, Yiyun Chen, Daojun Ye, and Haotian Yang are with The Hong Kong University of Science and Technology (Guangzhou), Guangzhou, China.
  \item Weikai Yang is with the Data Science and Analytics Thrust, Information Hub, The Hong Kong University of Science and Technology (Guangzhou), Guangzhou, Guangdong, China.
}

\abstract{%
Analysts often need to identify images that jointly satisfy multiple visual conditions, such as a crossroads with traffic lights at dusk, for model diagnosis, dataset curation, and targeted training.
Embedding-based retrieval can rank the large gallery efficiently, but a visually dominant condition can obscure weaker conditions, and a single similarity score does not enforce the required conjunction.
Visual question answering (VQA) can explicitly verify conditions, yet exhaustively applying it to the full gallery is costly, especially when analysts refine their query.
These limitations motivate keeping humans in the loop at the attribute level, where analysts can quickly build evidence for each condition and reuse it when the request changes.
We therefore present ProbeScout, a visual analytics system that supports this loop.
It first builds composable attribute probes from sparse VQA labels and fuses them into a conjunction-aware initial ranking.
Coordinated views support rapid screening, near-miss diagnosis, and on-the-fly subset construction by filtering and combining these probe outputs.
Analysts provide lightweight attribute- and query-level feedback, which drives staged refinement of fusion weights while keeping the probes fixed.
These verified attributes can be reused for future queries.
We evaluate ProbeScout on 17 retrieval tasks across three datasets, showing improved retrieval over embedding baselines. A separate 10-task comparison achieves higher task-macro AP and F1 than exhaustive VQA while labeling at most 2\% of the gallery images.
Two case studies further demonstrate how ProbeScout supports interactive analysis and refinement in realistic workflows.
}


\usepackage{amsmath}
\usepackage{amssymb}
\usepackage{algorithm}
\usepackage{algorithmic}
\usepackage{booktabs}
\AtEndPreamble{\DeclareCaptionLabelFormat{casepanel}{Figure~\thefigure.#2}}
\usepackage{mathptmx}
\usepackage{placeins}
\graphicspath{{figs/}{figures/}{pictures/}{images/}{./}}

\teaser{
  \centering
  \includegraphics[width=.90\linewidth,alt={ProbeScout interface annotated A through I: query specification, hierarchical evidence, visual embedding exploration, ranked gallery, diagnostic filters, selected image analysis, set overlap, human feedback, and refinement and validation.}]{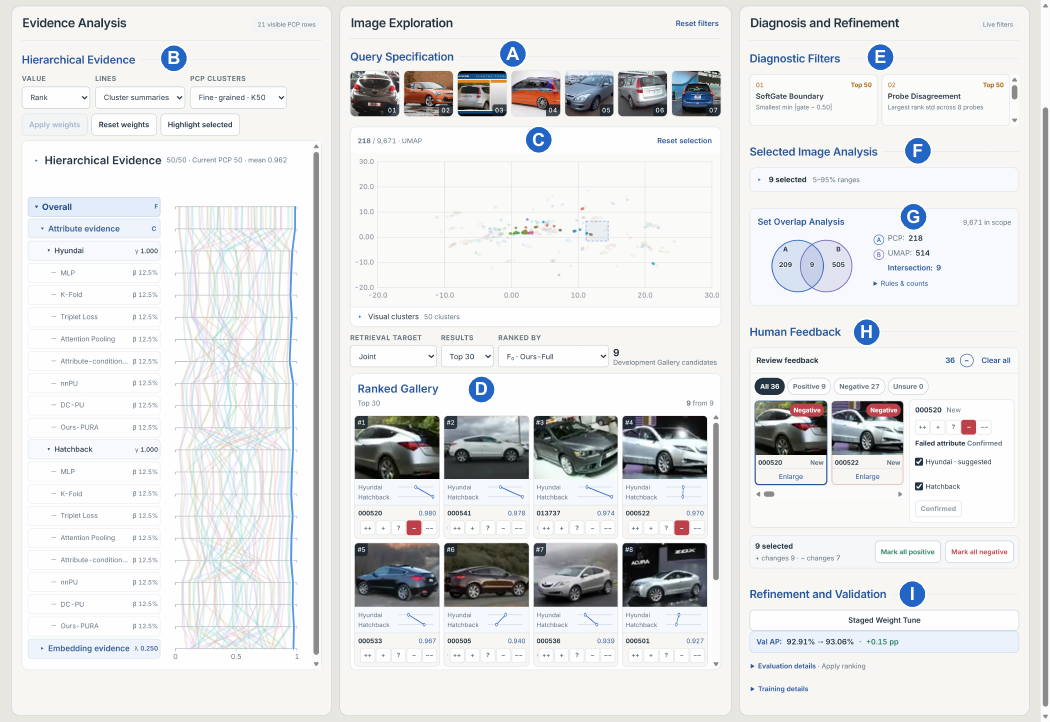}
  \caption{ProbeScout supports attribute-guided image retrieval over reusable attribute evidence.
  Analysts specify query conditions (A) and inspect their support across attribute probes (B). Linked embedding- and image-level views (C–D) reveal retrieval patterns and corresponding visual content, while diagnostic and refinement tools (E–I) support subset identification, feedback, and validation. Shared image identities enable seamless cross-view inspection.
  }
  \label{fig:interface}
}

\begin{document}

\firstsection{Introduction}
\maketitle

From a data-centric perspective, improving a vision model often depends on the subsets an analyst can extract from a large collection: training slices, failure cases, and evaluation groups~\cite{bertucci2022dendromap,gou2021vatld,zhang2023sliceteller}.
The useful subset is usually a conjunction of visual conditions rather than images that merely look similar, such as Hyundai hatchbacks or images of sheep being sheared.
Recovering such conjunctions from a large gallery is therefore a practical retrieval problem.
More importantly, this process is rarely a one-shot retrieval task.
Analysts inspect candidate images, discover unexpected cases, adjust conditions, and form different subsets as their understanding of the collection evolves.
Supporting this iterative screening process requires more than a single ranked list.

Existing automated methods address only parts of this workflow.
A common first idea is to embed images and conditions in a shared space and rank the gallery by similarity~\cite{yao2022filip,lim2026clay}.
While the ranking is very efficient, pooling several conditions into one score allows a visually dominant condition to mask weaker conditions.
The resulting ranking also provides little indication of which condition is satisfied or violated, making near misses difficult to diagnose and the ranking difficult to steer.
A second idea is to use visual question answering (VQA) to verify whether candidate images satisfy query conditions~\cite{feng2025vqa4cir}. Extending such verification to the full gallery, however, would incur substantial cost.
It becomes particularly inefficient during interactive exploration, where analysts may repeatedly add, remove, or replace conditions.
Human-in-the-loop visual analytics systems have explored ranking inspection, prompt refinement, and model or label validation~\cite{yang2024foundation,park2022vant,xuan2025attributionscanner,xuan2025vista}.
However, user judgments are commonly tied to the current query or item rather than retained as reusable evidence that can support subsequent exploration.

We therefore investigate an \emph{attribute-centric} workflow, where each visual condition becomes a persistent unit of both machine inference and human feedback.
This representation offers several advantages for visual analytics.
First, decomposing an aggregate score into per-condition evidence lets analysts inspect the basis for each judgment individually, rather than interpreting an opaque combined score.
Second, attribute-level judgments let analysts review images quickly and in batches, and feedback on ambiguous or borderline cases can refine how the probes for a condition are combined.
Third, attribute scores provide an expressive space for exploration, where analysts can filter, combine, and compare conditions to construct subsets that were not necessarily specified in the original query.
Finally, once an attribute has been verified, its evidence persists and can be recombined with other attributes as the analytical question evolves, so revisiting an earlier condition does not require re-examining the underlying images.

To support this workflow, we present \emph{ProbeScout}, a human-in-the-loop visual analytics system for attribute-guided exploration of large image collections.
The system first builds composable attribute probes from sparse, actively acquired VQA labels and preserves their outputs as separate, reusable evidence before combining them into a conjunction-aware ranking.
Coordinated views then let analysts screen the gallery, inspect which conditions hold, and freely construct subsets of interest from attribute scores.
Analysts can confirm or reject selected images at the level of a single attribute or of the full query.
The resulting attribute and query supervision drives staged optimization of fusion weights over frozen probes, and the revised ranking returns to the same views for further inspection.
Operating on stored probe outputs avoids retraining the attribute models at every interaction and keeps refinement lightweight.
As exploration proceeds, analysts can recombine existing attributes without repeating gallery-wide verification, while new attributes can be added through probe construction.
In this way, ProbeScout couples efficient machine-assisted ranking with lightweight human judgment and flexible subset exploration.

Our main evaluation comprises 17 retrieval tasks across three datasets:
seven from Stanford Cars, eight from HICO-DET, and two from CelebA.
Time-bounded feedback comparisons examine the additional value of coordinated analysis beyond the initial retrieval model. We also present two illustrative cases selected from repeated trials with researchers from AI and non-AI fields, covering human--object interaction retrieval and fine-grained vehicle retrieval.

The main contributions of this work are:

\begin{itemize}[nosep]
  \item We present \emph{ProbeScout}, an integrated visual analytics
  system that supports fine-grained multi-attribute image retrieval
  from textual conditions, query images, or both, combining
  reusable multi-probe ranking with human-guided analysis and
  refinement.

  \item We develop a label-efficient multi-probe retrieval method
  that actively acquires sparse VQA supervision, exploits
  unlabeled gallery images through positive--unlabeled pairwise
  learning, preserves separate evidence for individual attributes,
  and reuses learned probes across changing query combinations.

  \item We design a coordinated Scout workflow that combines
  hierarchical parallel coordinates, rank-profile clusters,
  visual-embedding clustering, overlap analysis, and linked image
  inspection to support gallery exploration, subset construction,
  feedback-image selection, and staged weight refinement at the attribute
  and full-query levels while keeping the probes fixed.

\end{itemize}

\section{Related Work}
\label{sec:related-work}

\subsection{Multi-Attribute and Composed Image Retrieval}
Composed image retrieval combines a reference image with a textual modification, through learned composition as in TIRG~\cite{vo2019composing} or zero-shot mappings as in Pic2Word~\cite{saito2023pic2word} and SEARLE~\cite{baldrati2023searle}. These mappings build on shared image--text representations such as CLIP~\cite{radford2021clip}. Reason-before-Retrieve instead uses multimodal reasoning to generate a target description without task-specific training~\cite{tang2025reasonbeforeretrieve}. However, evaluations of vision--language models reveal difficulties with attribute binding and relational distinctions~\cite{yuksekgonul2023bags}. Relative attributes~\cite{parikh2011relative} and WhittleSearch~\cite{kovashka2012whittlesearch} provide interpretable conditions for narrowing search. ProbeScout retains condition-level evidence for analysts to inspect, combine, and refine while building subsets beyond a single ranked list.

\subsection{Visual Exploration and Subset Construction}
Visual analytics has long supported model understanding and diagnosis by connecting model structures, learned representations, and individual instances. CNNVis helps experts understand, diagnose, and refine deep CNNs by visualizing neurons and their interactions \cite{liu2017towards}; ActiVis links model and instance views~\cite{Kahng2018ActiVis}; VAC-CNN compares convolutional networks~\cite{xuan2022vaccnn}; and Manifold relates model outputs to input characteristics without requiring model internals~\cite{zhang2019manifold}. ConceptExplainer organizes model explanations around discovered concepts~\cite{huang2023conceptexplainer}, while caption-guided image exploration connects semantic descriptions, visual features, and user steering~\cite{li2024imageexploration}. DendroMap supports hierarchical exploration of image collections~\cite{bertucci2022dendromap}, and InfoCIR combines composed retrieval with embedding exploration and explanations~\cite{dravilas2026infocir}. ProbeScout connects these concerns through named attribute evidence, rank-profile clusters, and visual-embedding clustering. Intersections and diagnostic filters construct image subsets whose evidence and selection provenance remain inspectable.

\subsection{Verification and Human-Guided Refinement}
Reusable concepts already support human-guided learning and search. CueFlik learns visual-concept rules from positive and negative examples for reranking subsequent image results~\cite{fogarty2008cueflik}; Visual Concept Programming lets experts compose visual concepts into labeling functions~\cite{Hoque2023VisualConceptProgramming}. FSLDiagnotor supports inspection and adjustment of base learners, ensemble weights, and representative examples for few-shot classification~\cite{Yang2022DiagnosingFewShot}. ProbeScout instead preserves a multi-attribute conjunction and refines attribute- and query-level fusion over frozen probes.

VQA4CIR checks candidate consistency through visual questions~\cite{feng2025vqa4cir}. Active learning addresses informative acquisition~\cite{tong2001support}, with core-set approaches emphasizing representativeness~\cite{sener2018coreset}. For label quality, Confident Learning estimates label errors~\cite{northcutt2021confidentlearning}, classifier-guided visual correction supports manual verification~\cite{baeuerle2020labelcorrection}, and Reweighter adjusts training-example weights~\cite{yang2024reweighting}. VISTA extends visual inspection to foundation-model-generated labels~\cite{xuan2025vista}. Building on relevance feedback~\cite{rui1998relevance} and semantic interaction~\cite{Endert2012SemanticInteraction}, ProbeScout links visual subset selection to explicit query and failed-attribute judgments. These judgments correct weak supervision and drive staged fusion-weight updates, rather than retraining the probes during interaction.

\section{Problem Analysis and Design Requirements}
\label{sec:problem}

\begin{figure*}[!t]
  \centering
  \includegraphics[width=\textwidth,alt={ProbeScout workflow: query specification, attribute and embedding evidence, visual analysis and subset selection, and feedback-driven refinement.}]{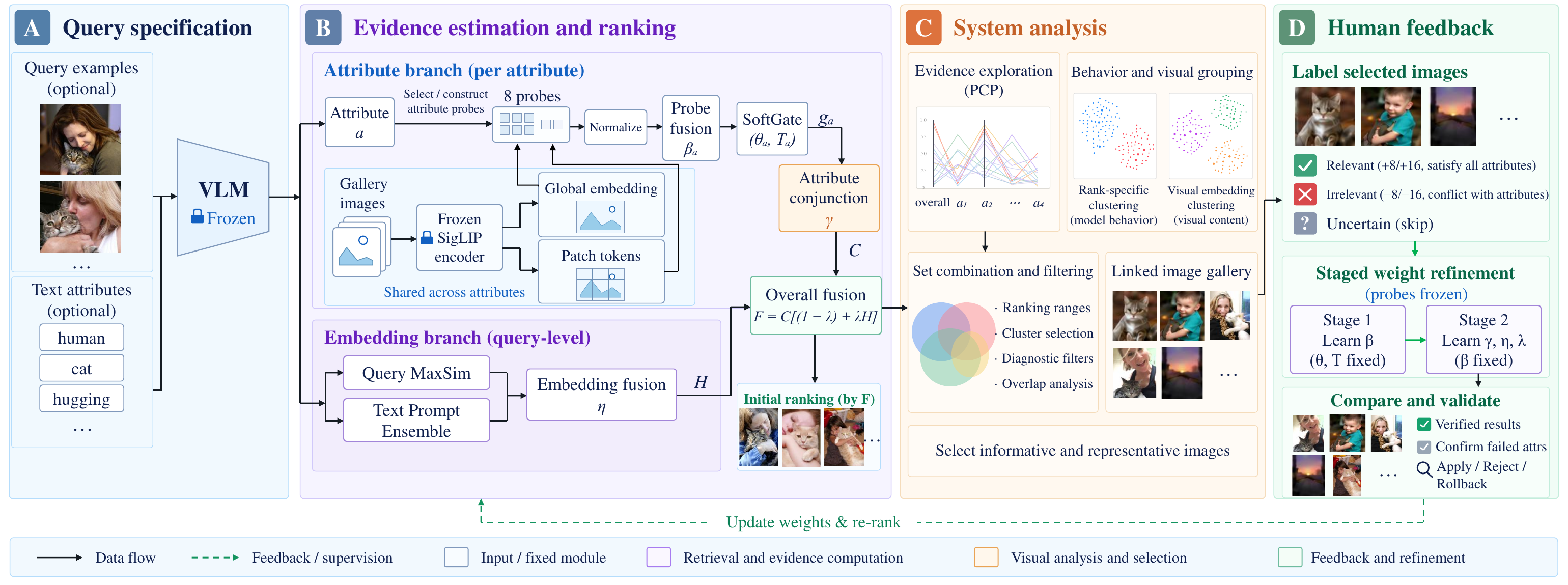}
  \caption{The ProbeScout workflow. Analysts specify a query from optional reference images and text and confirm its attributes (A). ProbeScout estimates reusable attribute evidence together with query-level embedding evidence and combines them into an initial ranking (B). Coordinated views expose the resulting evidence and retrieval behavior, supporting visual exploration, subset construction, and the selection of informative images (C). Analyst feedback on selected images is used to refine the ranking (D), forming an iterative analysis and refinement loop.}
  \label{fig:workflow-candidates}
\end{figure*}

Following the layered reasoning of the nested model for visualization design~\cite{munzner2009nested}, we characterize the domain challenges, derive the analytical tasks they entail, and translate these tasks into design requirements.

\subsection{Problem Context and Challenges}
ProbeScout targets analysts who work with large image collections for dataset curation and model diagnosis.
These analysts can assess whether an image satisfies specified visual conditions and recognize groups with shared visual content, but they should not be required to understand the underlying model architecture or manually tune retrieval and fusion parameters.
Multi-attribute retrieval is inherently conjunctive: an image is relevant only when it satisfies all required conditions.
Analysts therefore need to inspect retrieval results, diagnose recurring errors, construct informative subsets, and provide targeted feedback.
As retrieval intent evolves, they may also revise or recombine query conditions and reassess the resulting rankings.
These activities give rise to four recurring challenges:

\textbf{C1: Aggregated rankings obscure condition-level problems.} When several necessary conditions are compressed into a single score, analysts cannot readily determine the support for each condition or identify the condition that limits an image's rank. A negative judgment on an image indicates only that at least one condition is unmet, providing limited evidence about the source of the retrieval error.

\textbf{C2: Rank truncation creates inspection blind spots.} Leading results primarily expose images already prioritized by the model. Relevant images or entire visual groups that are incorrectly ranked low may fall outside the displayed top-$K$ results, where $K$ is the number of results shown, and remain difficult to discover through item-by-item inspection. Restricting inspection to leading results can systematically hide important false negatives and underrecognized visual patterns.

\textbf{C3: Semantically coherent error groups are difficult to identify.} False positives and false negatives for the same attribute may arise from different visual confusions. Selection based only on ranking position or appearance similarity may still mix distinct cases. Analysts need to combine and verify these local cues to construct coherent diagnostic subsets and select representative, nonredundant examples within a limited inspection budget.

\textbf{C4: Query evolution increases repeated verification costs.} As conditions are added, removed, or recombined, rankings and judgments associated only with the complete query may become difficult to reuse. Analysts may repeatedly verify the same visual evidence across related queries, while feedback obtained for one conjunction of conditions may provide limited support for reasoning about another. Such repetition increases the cost of iterative retrieval and refinement.

\subsection{Analytical Tasks}
Inspired by established distinctions between analytical goals and operations in visualization task abstraction~\cite{brehmer2013typology}, we organize the workflow into five connected analytical tasks.

\textbf{T1: Define and revise retrieval conditions.} Specify the visual conditions that retrieved images must satisfy using text, reference images, or both, and revise these conditions as the retrieval intent evolves by adding, removing, or recombining conditions.

\textbf{T2: Diagnose ranking and condition-level problems.} Examine how individual conditions contribute to the overall ranking, identify conditions that systematically limit retrieval quality, and determine whether individual ranking errors indicate recurring failure patterns that warrant further investigation.

\textbf{T3: Construct diagnostically informative image subsets.} Locate groups of potentially relevant or problematic images by jointly considering retrieval behavior and visual content. Compare candidate groups, inspect their common characteristics, and construct focused subsets for subsequent analysis or feedback. 

\textbf{T4: Verify relevance and attribute-level failures.} Judge whether selected images satisfy the complete query and confirm one or more failed attributes for negative examples, producing supervision for retrieval refinement. 

\textbf{T5: Validate improvements and decide subsequent actions.} Compare retrieval results before and after refinement to determine whether an update corrects the targeted failure patterns, preserves previously relevant images, and avoids introducing new errors. 

Together, these tasks connect intent, diagnosis, and subset construction with feedback and outcome validation. Analysts may stop with a reviewed subset or revisit earlier tasks as their analysis evolves. 

\subsection{Design Requirements}

Based on the above challenges and analytical tasks, we derive four design requirements for supporting efficient and interpretable multi-attribute retrieval analysis.

\textbf{R1: Preserve decomposable and reusable attribute evidence (T1, T2).} The system should retain each required attribute and its supporting evidence so that analysts can trace the overall ranking to individual conditions. Attribute definitions, evidence provenance, and query-local configurations should be managed separately, allowing compatible evidence to be reused when attributes are recombined across related queries.

\textbf{R2: Connect retrieval behavior with visual content (T2, T3).} The system should enable analysts to compare patterns in model responses with patterns in image content. It should support transitions from individual images to groups exhibiting similar retrieval behavior, as well as from group-level patterns back to concrete examples, helping analysts assess whether shared model responses correspond to recognizable visual characteristics or common failure modes.

\textbf{R3: Support composable and traceable subset construction (T3).} The system should allow analysts to progressively construct diagnostic subsets by combining rank ranges, attribute-evidence patterns, visual groups, and other diagnostic cues. Subset construction should extend beyond the displayed top-$K$ results so that analysts can investigate potentially relevant but under-ranked regions of the collection. 

\textbf{R4: Support low-burden feedback and verifiable refinement (T4, T5).} The system should let analysts directly judge complete-query relevance and failed attributes for selected images and translate these judgments into the corresponding supervision, without requiring manual parameter adjustment. It should also provide before-and-after comparisons of retrieval results and validation metrics so that analysts can inspect intended improvements and potential adverse effects. These capabilities are consistent with established human--AI interaction principles emphasizing correction, dismissal, and user control~\cite{amershi2019guidelines}.

\section{ProbeScout System Overview}
\label{sec:overview}

The tasks and requirements in \cref{sec:problem} call for decomposable attribute evidence, inspection beyond the displayed top-$K$, subset construction that links ranking behavior to visual content, and low-burden feedback with verifiable updates.
ProbeScout supports this work through the loop in \cref{fig:workflow-candidates}: construct reusable attribute evidence, inspect and subset the gallery, then refine the ranking with two-level feedback.
\Cref{tab:design-mapping} maps T1--T5 and R1--R4 onto the components of this loop.

\begin{table*}[tb]
\caption{Task--requirement--component mapping. Each component supports a decision within the analytical loop; shared selection connects the decisions across views.}
\label{tab:design-mapping}
\centering\normalsize
\begin{tabular}{@{}p{.16\textwidth}p{.10\textwidth}p{.28\textwidth}p{.38\textwidth}@{}}
\toprule
Analytical task & Requirements & Components & Supported decision \\
\midrule
T1: Define conditions & R1 & Attribute confirmation and evidence registry & Reuse a condition, construct a missing attribute, or revise the active conjunction. \\
T2: Diagnose ranking issues & R1, R2 & Hierarchical parallel coordinates and image profiles & Identify weak attributes and recurring ranking errors. \\
T3: Locate candidate groups & R2, R3 & Rank-profile clusters and visual-embedding clustering & Compare visually similar groups with groups sharing ranks across the scoring hierarchy. \\
T3: Construct image subsets & R3 & Diagnostic filters, overlap summary, and linked gallery & Combine selection cues and inspect subsets for analysis or representative, nonredundant feedback. \\
T4: Provide supervision & R4 & Image-level feedback and failed-attribute confirmation & Confirm query relevance and failed attributes to guide staged weight refinement. \\
T5: Validate improvements & R4 & Before/after metrics, shared ranked gallery, and rollback & Inspect corrections and new ranking errors, then accept, reject, or roll back the update. \\
\bottomrule
\end{tabular}
\end{table*}

Analysts specify a query $q$ with reference images, textual conditions, or both.
For image input, a vision--language model proposes candidate attributes, which analysts confirm, remove, or supplement.
The system then loads compatible registered probes or constructs probes for missing attributes, so each required condition remains named, reusable evidence (T1, R1).
Retrieval combines two branches.
The attribute branch fuses each attribute's probe outputs into per-attribute support and multiplies these supports into a conjunction score, so an image ranks highly only when every required attribute holds.
The query-level embedding branch scores each gallery image against the request as a whole and modulates this conjunction rather than replacing it (\cref{sec:calibration}).
Images are default-ranked by the combined score for quick inspection, while retained per-attribute support lets analysts see which conditions hold rather than reading only a pooled rank (R1).

Coordinated views then turn this evidence into inspectable subsets.
Hierarchical parallel coordinates and image profiles expose per-condition support, so ranking errors can be traced to weak or conflicting conditions (T2).
Rank-profile clusters and visual-embedding clustering contrast ranking behavior with visual groups, helping analysts tell a shared failure pattern from a visually mixed set (T3, R2).
Diagnostic filters, an overlap summary, and a linked gallery combine these cues so that subset construction is not limited to the displayed top-$K$ (T3, R3).
Analysts may keep a reviewed subset as an analysis output, or provide query-level and attribute-level judgments on selected images (T4, R4).
Requested refinement updates fusion weights over frozen probes and returns a candidate ranking for before/after comparison, validation metrics, and rollback (T5).
As the query evolves, verified attributes can be recombined without repeating gallery-wide verification.

\section{Reusable Attribute Evidence}
\label{sec:evidence}
This section describes how ProbeScout turns a query into the scores that the views display, and how the resulting attribute evidence is stored for reuse.

\subsection{Scoring Overview}
\label{sec:scoring-overview}
Scoring proceeds in four steps.
The query is first resolved into a set of named attributes that a relevant image must satisfy (\cref{sec:query-attributes}).
Each attribute then receives a bank of lightweight probes, fitted from sparse VQA judgments over frozen visual features (\cref{sec:sparse-verification}).
An attribute's probe scores are normalized to a common scale, fused into one support value, and gated to a graded degree of satisfaction (\cref{sec:calibration}).
The gated supports finally form a soft conjunction score, which complete-query embedding evidence modulates into the ranking score (\cref{eq:initial-ranking}).

Two properties of this decomposition support the analysis loop.
First, every intermediate quantity remains addressable for an image: one score per probe, one support value per attribute, a conjunction score, and a final score, so a low rank can be traced to the attribute that limits it (R1).
Second, a probe is fitted for one named attribute rather than for a complete query, so the same evidence serves later queries that add, remove, or recombine conditions (\cref{sec:evidence-reuse}).
Analyst feedback adjusts only the fusion parameters introduced below and leaves the probes frozen (\cref{sec:refinement}).

\subsection{Query and Attribute Definition}
\label{sec:query-attributes}
Let $\mathcal X=\{x_i\}_{i=1}^{N}$ contain $N$ gallery images indexed by $i$, and let $\mathcal A_q$ be the nonempty set of canonical attributes confirmed for query $q$, with $a$ indexing an attribute.
For reference images, Qwen-VL-Max~\cite{bai2023qwenvl,alibaba2026qwenvlapi} first describes visible content in each image, then summarizes these descriptions into atomic attributes suitable for binary labeling.
A rule-based filter retains up to six candidates that the model reports as present in all reference images, with high verifiability and low or medium subjectivity.
Analysts confirm, remove, or supplement these candidates, excluding incidental image content.
The confirmed union of image-derived and textual attributes forms $\mathcal A_q$.

\subsection{Attribute Probes from Sparse Judgments}
\label{sec:sparse-verification}
Every probe reads the same frozen SigLIP features~\cite{zhai2023siglip}, so adding an attribute trains a small scorer rather than the backbone.
A probe is a lightweight attribute scorer; $m$ indexes $M=8$ configurations that differ in architecture, loss, or training strategy, six reading the pooled image vector and two reading patch tokens.
Probe $m$ for attribute $a$ maps image $x_i$ to a probability $s_{iam}$ through the logistic sigmoid $\sigma$.
Feature dimensions and training settings are given in Appendix~\ref{appendix:implementation}.

\paragraph{Iterative VQA acquisition.}
Initialization selects 100 distinct images using query-prototype similarity, attribute-text coverage when available, and embedding diversity.
Diversity-aware selection uses an image-embedding adaptation of maximal marginal relevance~\cite{carbonell1998mmr}.
A fixed VQA annotator supplies binary judgments for the required attributes.
An MLP and a positive--unlabeled ranking probe form an acquisition committee, refitted on accumulated fitting labels before each round.
Its disagreement cue follows the query-by-committee principle~\cite{seung1992qbc}.
Each round selects 40 additional fitting images and 10 disjoint audit images, combining positive discovery, conjunction-boundary proximity, committee disagreement, and diversity with class-balance guards.
This lightweight committee acquires supervision for the final eight-configuration evidence set.

Audit predictions are recorded before labeling, and audit labels never enter fitting, checkpoint selection, or calibration.
Both fitting and audit images count toward the logical budget.
Acquisition ends at the configured budget or round limit; adaptive runs also use audit-positive yield and attribute coverage to determine early stopping or bounded extension.
Fixed-budget comparisons disable adaptive stopping and use prescribed trajectory prefixes.
The 50-image setting uses a deterministic subset of the 100-image initialization, stratified by acquisition role.
Fitting uses the original non-audit VQA supervision under the data protocol in \cref{sec:evaluation-protocol}.

\paragraph{Eight probe configurations.}

For each attribute, we construct eight lightweight probes on frozen visual features, spanning supervised learning, positive--unlabeled (PU) classification and ranking, and patch-based attention. Their separate outputs support attribute-level fusion and inspection of probe disagreement. All eight configurations are retained, with checkpoints selected on the fixed validation set (Val).
Appendix~\ref{appendix:probe-configurations} details the configurations.

\subsection{Attribute Support and Conjunction Ranking}
\label{sec:calibration}
\paragraph{Attribute support.}
Probe probabilities are trained separately and are therefore not directly comparable across configurations, so ProbeScout normalizes each probe to $\widetilde s_{iam}\in[0,1]$ before fusion using min-max normalization.
The normalized scores of an attribute are then combined by a probe-weight vector $\beta_a=(\beta_{am})_{m=1}^{M}$, with $\beta_{am}\geq0$ and $\sum_{m=1}^{M}\beta_{am}=1$, so that configurations better suited to that attribute can carry more of its evidence.
A soft gate finally converts the fused score into a graded degree of satisfaction, giving the fused attribute score $u_{ia}$ and the SoftGate output $g_{ia}$:
\begin{equation}
u_{ia}=\sum_{m=1}^{M}\beta_{am}\widetilde s_{iam},
\qquad g_{ia}=\sigma\!\left(\frac{u_{ia}-\theta_a}{T_a}\right).
\label{eq:attribute-gate}
\end{equation}
A hard decision at $\theta_a$ would discard how far an image sits from the threshold, whereas the gate retains this margin, keeping clear matches, near misses, and clear failures distinguishable once attributes are combined.
Both gate parameters are selected under uniform probe weights on the fixed Validation set defined by the same protocol in \cref{sec:evaluation-protocol}.

\paragraph{Soft conjunction and ranking score.}
The gated supports must now become one ranking score that still reflects the required conjunction of attributes.
ProbeScout therefore multiplies them into a conjunction score $C_i$, weighted by attribute exponents $\gamma=(\gamma_a)_{a\in\mathcal A_q}$ with $\gamma_a>0$.
The product implements a \emph{soft conjunction}: weak support for any required attribute lowers $C_i$ regardless of how strongly the remaining attributes hold.
A query can also carry cues that no single attribute captures, so two query-level scorers, Query MaxSim and Text Prompt Ensemble, compare each image with the complete request.
Their scores $E_{i1}$ and $E_{i2}$, min--max-normalized on the same DG rows, are combined by embedding weights $\eta=(\eta_1,\eta_2)$, with $\eta_1,\eta_2\geq0$ and $\eta_1+\eta_2=1$, into complete-query evidence $H_i$.
This evidence modulates rather than replaces the conjunction, giving the overall score $F_i$:
\begin{equation}
\begin{aligned}
C_i&=\prod_{a\in\mathcal A_q}g_{ia}^{\gamma_a},
& H_i&=\eta_1E_{i1}+\eta_2E_{i2},\\
F_i&=C_i\big[(1-\lambda)+\lambda H_i\big],
\end{aligned}
\label{eq:initial-ranking}
\end{equation}
Here $\lambda\in[0,1]$ sets how far complete-query evidence may scale the conjunction; because this factor multiplies $C_i$, a high overall similarity cannot compensate for an attribute that fails.
The initial configuration $F_0$ combines these selected gate parameters with $\beta_{am}=1/8$, $\gamma_a=1$, $\eta_1=\eta_2=1/2$, and $\lambda=\lambda_0=0.25$, before human-feedback refinement.
The product is computed in the log domain.
Sorting images by $F_i$ in descending order defines the initial ranking $\pi_q^0$.

\subsection{Evidence Registration and Reuse}
\label{sec:evidence-reuse}
Each record stores the canonical attribute, feature and backbone identity, configuration, checkpoint or fallback, supervision provenance, and gallery scores.
Reusable attribute evidence and calibration records are separate from query-local fusion parameters.
Compatible registered attributes can be recombined without retraining their probes, while missing attributes require construction from representative images.
Label-free recomposition reuses stored attribute calibration, whereas selecting a new query-local gate configuration requires complete conjunction labels on the validation set.

\section{Multi-View Subset Construction}
\label{sec:visual-analytics}

\subsection{Interface and Coordinated Views}
\label{sec:interface-views}

As shown in \cref{fig:interface}, ProbeScout organizes the workflow into three coordinated regions: evidence analysis (B), image exploration (A, C, D), and diagnosis and refinement (E--I). Analysts begin with the query and ranked results (A, D), inspect attribute-level support in the evidence view (B), and relate retrieval behavior to visual structure through the embedding view (C). Diagnostic views (E--G) help identify and inspect candidate groups, which can then be judged in the feedback panel (H) and reviewed after refinement in the validation panel (I). Shared image identities coordinate these interactions across views~\cite{roberts2007coordinated}, including selected images beyond the displayed top-$K$. Exploration itself does not alter the model, and only explicit judgments contribute to refinement.

\subsection{Attribute Evidence and PCP Brushing}
\label{sec:hierarchical-diagnosis}

Parallel coordinates support multivariate comparison~\cite{inselberg1990parallel}, and hierarchical extensions organize large collections at multiple levels~\cite{fua1999hierarchical}. Our hierarchy follows the scoring structure rather than a hierarchy of data clusters (R1). The overall score expands into conjunction and complete-query evidence; the former expands into attribute gates and their eight probes, and the latter into Query MaxSim and Text Prompt Ensemble. Rank axes compare relative ordering, while score axes show the corresponding model outputs, with normalized evidence at probe and embedding leaves. A gate output near $0.5$ indicates proximity to its transition; the threshold $\theta_a$ acts on the pre-gate fused score $u_{ia}$, not on the displayed gate value. Brushing several axes specifies an evidence pattern, and progressive expansion reveals whether an aggregate conceals disagreement among its components.

In the hierarchical evidence view (\cref{fig:interface} (B)), a brushed pattern defines a subset that is linked to the embedding view (C) and gallery (D). Conversely, selecting an image exposes its evidence profile in (B), allowing analysts to locate other images with similar attribute-level behavior. For a Hyundai-hatchback query, for example, brushing strong Hyundai evidence but weak hatchback evidence isolates near misses such as Hyundai sedans or SUVs. Expanding the hatchback attribute reveals whether its probes consistently reject these images or disagree.

\subsection{Rank-Profile Clusters and Visual-Embedding Clustering}
\label{sec:contrasting-spaces}

\emph{Rank-profile clusters} group images by normalized ranks across the scoring hierarchy, including the overall score, conjunction, attribute gates, probes, and embedding branch. Their clustering basis remains rank-based when PCP switches to score display. In contrast, \emph{visual-embedding clustering} groups images in normalized frozen feature space and displays them in a two-dimensional projection. Selecting a visual group in \cref{fig:interface}(C) highlights its members in the linked gallery (D).

The two spaces expose complementary structure (R2): rank-profile clusters reveal images treated similarly by the retrieval model, whereas visual-embedding clusters reveal images that look similar. Comparing them can expose visually coherent groups receiving inconsistent evidence, or visually diverse images sharing the same ranking behavior. The visual view supports PCA or UMAP~\cite{mcinnes2018umap}; linked image inspection helps verify projected neighborhoods because dimensionality reduction may distort high-dimensional relationships~\cite{jeon2025backstage}. Clustering and projection settings are summarized in \cref{sec:clustering-settings}.

\subsection{Filtering and Set Combination}
\label{sec:candidates}

\paragraph{Combine selection conditions.}
Each active PCP range, selected rank-profile cluster, selected visual-embedding cluster, and projection brush defines a set of image identities in the current scope. One cluster can be active in each space, and the intersection of active conditions forms the candidate pool. In \cref{fig:interface}(F), Selected Image Analysis provides rank and score context for the currently inspected image, while Set Overlap Analysis (G) summarizes intersections among active selections. The Venn-based summary reports individual-condition and shared-membership counts (R3), two central set-analysis concerns~\cite{alsallakh2016setvisualization}. For up to three conditions, it shows exact exclusive-region counts with schematic circle areas; for more conditions, it shows the all-condition intersection and cumulative counts. 
Analysts broaden or narrow the pool by adjusting the active rules and reviewing the updated gallery.

\paragraph{Apply diagnostic filters.}
The Diagnostic Filters panel (\cref{fig:interface}E) provides four targeted entry points into these candidate sets. They use a pre-exported diagnostic snapshot held fixed across ranking configurations. \emph{SoftGate boundary} prioritizes the smallest distance to gate output $0.5$ across active attributes. \emph{Probe disagreement} measures the standard deviation of the eight probes' rank percentiles. \emph{Individual-probe rescue} finds images ranked highly by a selected probe but poorly by cached fusion. \emph{Prototype--ranking mismatch} finds high prototype-similarity ranks accompanied by low ranks under a selected comparison scorer. One diagnostic filter can be active at a time; it narrows the intersection to at most 50 inspection candidates and remains separate from the condition-overlap summary. Candidates remain linked to the ranked gallery (D), where analysts select representative, nonredundant examples for feedback. Changing a brush revises the selection, whereas changing a required attribute revises the query.

\subsection{Staged Weight Refinement}
\label{sec:refinement}

\paragraph{Feedback supervision.}
Analysts assign positive, negative, or uncertain judgments on gallery cards (D), with ordinary or strong confidence for positive and negative feedback; the Human Feedback panel (\cref{fig:interface}H) collects these judgments for review. Positive query judgments confirm all required attributes. For a negative example, the interface suggests a failed attribute based on the weakest initial gate, which analysts can confirm or revise. Only confirmed failures receive negative attribute labels; other attributes retain existing VQA labels or remain unknown (R4). Uncertain judgments add no feedback supervision and preserve existing VQA labels.

Refinement combines accumulated eligible DG feedback with the original VQA fitting supervision. Human judgments override corresponding original labels, and strong feedback receives greater weight than ordinary feedback. Supervision is balanced separately for each attribute and for the query. Attribute judgments retain their named conditions, while complete-query evidence and its fusion parameters remain query-specific.

\paragraph{Two-stage optimization.}
Both stages use the score model in \cref{eq:attribute-gate,eq:initial-ranking}, keeping the backbone, probes, normalization statistics, and SoftGate thresholds and temperatures frozen. Each run starts from the same initial fusion configuration using accumulated supervision. Stage 1 updates only the probe-fusion weights $\beta$ using attribute-level supervision. Stage 2 fixes $\beta$ and updates the attribute exponents $\gamma$, embedding weights $\eta$, and mixing weight $\lambda$ using query-level supervision. Both stages use class-balanced binary cross-entropy with regularization toward the initial configuration.

Because the fusion parameters are shared across gallery images, query-level supervision can change rankings beyond the feedback set without retraining the probes. In Stage 2, positive query examples supply gradients to all attribute exponents, whereas negative examples supply them only to exponents of failed attributes identified by VQA labels or confirmed by the analyst. Both positive and negative examples supply gradients to the embedding branch. The system retains the state with the lowest combined regularized training objective among the initial, completed Stage-1, and completed Stage-2 states, then reports the candidate ranking and fixed-Val metrics for analyst comparison. Appendix~\ref{appendix:optimization} specifies the supervision weights, objectives, constraints, and optimization settings.

\subsection{Ranking Validation and Rollback}
\label{sec:validation}

The Refinement and Validation panel (\cref{fig:interface}I) presents the candidate update together with before/after Val metrics and learned fusion parameters. Analysts can toggle between the initial and candidate configurations, with the ranked gallery (D) updating accordingly to compare numerical changes with concrete retrieval results. The linked evidence view (B) further shows how fixed probe evidence is reweighted for images whose ranks change substantially.

Applying a candidate activates its fusion; rollback restores the initial configuration and ranking (R4). Switching rankings clears PCP brushes and rank-profile cluster selections whose meanings depend on the scoring state, while confirmed judgments remain reviewable. Rejecting a fit therefore does not discard the feedback.
\section{Evaluation}
\label{sec:evaluation}
We organize the evaluation around retrieval effectiveness, supervision efficiency and reuse, feedback-driven improvement, and the value of coordinated visual analysis.

\subsection{Research Questions and Protocol}
\label{sec:evaluation-protocol}
The evaluation addresses four research questions:

\textbf{RQ1: Retrieval effectiveness.} Do attribute evidence and its fusion improve fine-grained, multi-attribute retrieval quality?

\textbf{RQ2: Supervision efficiency and reuse.} What retrieval quality can be achieved with a limited VQA budget, and can existing attributes support new conjunctions at low additional cost?

\textbf{RQ3: Feedback-driven improvement.} Can a small amount of query-level and attribute-level supervision improve held-out rankings, and how do the update mechanisms contribute?

\textbf{RQ4: Visual analysis value.} How do coordinated views help analysts diagnose error groups, select informative feedback images, and manage analysis effort?

\paragraph{Tasks and data roles.}
The main evaluation comprises 17 tasks from three datasets: Stanford Cars~\cite{krause2013cars} (7), HICO-DET~\cite{chao2018hicodet} (8), and CelebA~\cite{liu2015celeba} (2). These tasks were retained after reviewing the original attribute annotations, and the sampling and exclusion process is described in Appendix~\ref{appendix:task-scope}. Method comparisons, core component ablations, VQA-budget evaluation, and feedback comparisons use this same task set. Attribute reuse is evaluated on 12 new query combinations from these datasets. A separate comparison with exhaustive VQA uses 10 tasks because of the cost of full-gallery annotation (\cref{sec:exhaustive-vqa-comparison}). Each task defines a gallery, attributes, reference positives, and image-ID splits; assessment labels are distinct from VQA supervision.

Development Gallery (DG) supports inspection and feedback. A single fixed VQA Validation (Val) set contains approximately 20\% of the original VQA supervision, using split seed 0 and shared image identities across attributes. It supports probe-checkpoint selection, initial SoftGate parameter selection, classification-threshold selection, and analyst comparison of refinement candidates. Original VQA fitting rows comprise the acquired non-audit supervision. Human feedback and normalization use eligible DG rows; the original fitting rows need not all belong to the DG display mask. Val, Test, and query-reference images are excluded from these fitting, feedback, and normalization rows and from the unlabeled PU pool. Staged weight fitting and internal refinement-state selection use the training objective, while Test provides final held-out evaluation. Gallery metrics cover the full collection, including fitting and feedback images.

\paragraph{Models and training.}
We use a frozen SigLIP ViT-B/16 visual encoder~\cite{zhai2023siglip} and Qwen-VL-Max~\cite{alibaba2026qwenvlapi} for query-attribute extraction and candidate attribute labeling. Each attribute has eight probes trained with five random seeds. Probe checkpoints are selected by ROC-AUC on fixed Val. Refinement uses the staged updates in \cref{sec:refinement}, with probes frozen throughout. Model dimensions, training hyperparameters, and clustering and refinement settings are provided in Appendix~\ref{appendix:implementation}.

\paragraph{Budget accounting.}
We distinguish the VQA budget for constructing probes from the human-feedback budget for interactive refinement. The logical VQA budget counts acquired fitting and audit images per task. The reported labeling percentage is the number of VQA-labeled images divided by the task's gallery size, rather than the number of attribute judgments or API calls. For a gallery of 30,000--40,000 images, a 300-image budget corresponds to 0.75--1\% of the collection. Inspected images, query-level judgments, and image--attribute judgments are separate quantities because one image may yield several attribute labels. The VQA-budget comparison evaluates 50--300 images per task with adaptive stopping disabled. Acquisition schedules are detailed in Appendix~\ref{appendix:calibration-settings}.

\paragraph{Measures, comparisons, and statistics.}
The quantitative comparisons report equally weighted task-macro AP and F1, with Test assessing held-out retrieval quality and Gallery describing the full collection. Probe probabilities are averaged across five seeds within each attribute before conjunction and metric calculation. In the 17-task method comparison and component ablations, every scorer selects its final classification threshold $\tau$ to maximize F1 on the same original VQA Val query-level labels, then applies it to Test and Gallery.

For configurations with SoftGate, Val AP selects a threshold offset and temperature multiplier shared across attributes within each task, while fusion weights remain at their initial values. The resulting gate parameters $\theta_a$ and $T_a$ are part of $F_0$ and remain fixed during staged weight refinement. Fusion and Full ProbeScout in \cref{tab:probe-performance,tab:component-ablations} denote the same $F_0$ configuration and use identical parameters and scores. In the separate VQA-budget experiment, AP is averaged over five seeds within each task and then equally across tasks, and 95\% confidence intervals use paired task-bootstrap resampling~\cite{efron1979bootstrap}. Appendix~\ref{appendix:calibration-settings} gives the calibration grid and resampling settings.

\subsection{Quantitative Evaluation}
\label{sec:quantitative-evaluation}

\subsubsection{Initial Retrieval Performance}
\label{sec:initial-retrieval-evaluation}
\begin{table}[tb]
  \caption{Retrieval performance on the 17 evaluation tasks. Values are task-macro AP and F1 (\%). Bold and underlining indicate the best and second-best values, respectively.}
  \label{tab:probe-performance}
  \centering\normalsize
  \setlength{\tabcolsep}{3pt}
  \begin{tabular}{@{}lrrrr@{}}
    \toprule
    & \multicolumn{2}{c}{Test} & \multicolumn{2}{c}{Gallery} \\
    \cmidrule(lr){2-3}\cmidrule(l){4-5}
    Scorer & AP & F1 & AP & F1 \\
    \midrule
    Query MaxSim & 65.00 & 53.53 & 66.45 & 56.21 \\
    Text Prompt Ensemble & 65.67 & 54.72 & 63.89 & 56.13 \\
    CLAY & 50.60 & 43.05 & 51.10 & 45.23 \\
    MLP & 72.13 & 59.97 & 71.76 & 60.71 \\
    K-Fold & 73.68 & 64.61 & 73.92 & 64.77 \\
    Triplet Loss & 78.45 & 65.12 & 78.15 & 68.42 \\
    Attention Pooling & 79.46 & \underline{67.42} & \underline{79.22} & \underline{69.36} \\
    Attribute-conditioned Attention & 78.62 & 65.03 & 78.18 & 66.70 \\
    nnPU & 74.96 & 59.98 & 74.62 & 60.57 \\
    DC-PU & 73.58 & 59.75 & 73.51 & 60.84 \\
    PURA & \underline{79.85} & 66.75 & 79.16 & 68.24 \\
    \textbf{Fusion ($F_0$)} & \textbf{83.21} & \textbf{72.27} & \textbf{82.33} & \textbf{73.65} \\
    \bottomrule
  \end{tabular}
\end{table}

To address RQ1 before human feedback, \cref{tab:probe-performance} compares embedding baselines, individual probes, and initial fusion on the 17 evaluation tasks. Fusion reaches Test AP 83.21\% and Gallery AP 82.33\%. The strongest embedding baselines are Text Prompt Ensemble on Test (65.67\%) and Query MaxSim on Gallery (66.45\%), giving gains of 17.54 and 15.88 percentage points, respectively. The highest-AP individual probes are PURA on Test (79.85\%) and Attention Pooling on Gallery (79.22\%); Fusion improves upon them by 3.36 and 3.11 percentage points. Fusion also achieves the highest F1 on both Test and Gallery (72.27\% and 73.65\%, respectively). These results support combining probe evidence to provide a ranking for visual analysis.

\subsubsection{VQA Budget and Attribute Reuse}
\label{sec:budget-reuse-evaluation}
\begin{figure}[tb]
  \centering
  \includegraphics[width=0.8\columnwidth]{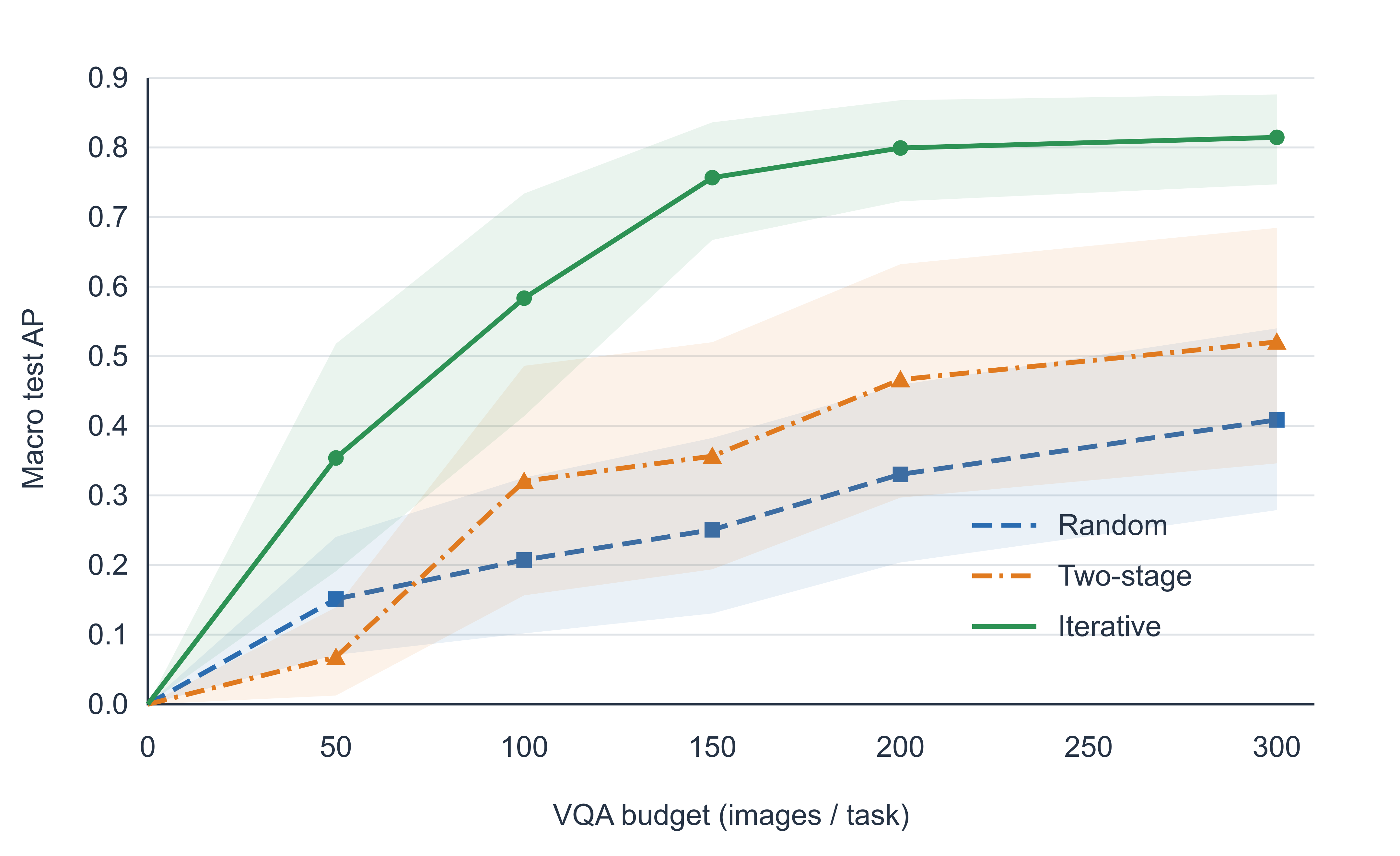}
  \caption{Comparison of Random, Two-stage, and Iterative VQA acquisition across the 17 evaluation tasks. Macro Test AP is plotted against the logical VQA budget (images per task). Shaded bands indicate 95\% task-bootstrap confidence intervals.}
  \label{fig:vqa-efficiency}
\end{figure}

\paragraph{VQA acquisition efficiency (RQ2).}
\Cref{fig:vqa-efficiency} shows Iterative acquisition across logical VQA budgets of 50--300 images per task on the 17 evaluation tasks. This experiment uses attribute-only fusion of the eight probes with SoftGate and a product across attributes. Macro Test AP increases from 35.38\% at 50 images to 75.66\% at 150, 79.92\% at 200, and 81.45\% at 300, with smaller gains beyond 150 images. These measurements characterize the supervision budget needed to construct the initial attribute evidence.

\begin{table}[tb]
  \caption{Attribute reuse on 12 new query combinations: task-macro Test AP and F1 (\%). F1 uses original attribute-training thresholds followed by AND. Complete results are in Appendix~\ref{appendix:reuse-results}.}
  \label{tab:probe-reuse}
  \centering\normalsize
  \setlength{\tabcolsep}{3pt}
  \begin{tabular}{@{}lrr@{}}
    \toprule
    Method & AP & F1 \\
    \midrule
    Attribute-text mean & 49.00 & 38.73 \\
    ProbeScout Reuse & \textbf{75.60} & \textbf{63.69} \\
    \bottomrule
  \end{tabular}
\end{table}

\paragraph{Attribute reuse (RQ2).}
\Cref{tab:probe-reuse} separately evaluates 12 new query combinations supported by the available evaluation records, using existing attribute probes without retraining. The attribute embedding baseline, Attribute-text mean, averages cosine similarities between image embeddings and the query attributes' SigLIP text embeddings. Limited overlap between source-task training sets makes joint calibration for new attribute combinations unreliable. We therefore compute F1 using the original attribute thresholds with a logical AND, without new query-level labels, and compute AP from the continuous ranking score. ProbeScout Reuse achieves macro Test AP of 75.60\% and F1 of 63.69\%, compared with 49.00\% and 38.73\% for Attribute-text mean. These results support compositional reuse of known attribute probes for new query combinations, although gains vary by task: for \emph{Jump skateboard}, Reuse has lower AP than Attribute-text mean (64.20\% versus 79.20\%). Complete per-query results appear in Appendix~\ref{appendix:reuse-results}.

\subsubsection{Label Efficiency Compared with Exhaustive VQA}
\label{sec:exhaustive-vqa-comparison}
\begin{table}[tb]
  \caption{Comparison with exhaustive VQA on 10 tasks: eight Stanford Cars tasks and two HICO-DET tasks. Values are equally weighted task-macro AP and F1 (\%). Bold indicates the best value in each column.}
  \label{tab:exhaustive-vqa}
  \centering\normalsize
  \setlength{\tabcolsep}{3pt}
  \begin{tabular}{@{}lrrrr@{}}
    \toprule
    & \multicolumn{2}{c}{Test} & \multicolumn{2}{c}{Gallery} \\
    \cmidrule(lr){2-3}\cmidrule(l){4-5}
    Method & AP & F1 & AP & F1 \\
    \midrule
    Query MaxSim & 77.81 & 61.46 & 78.87 & 62.13 \\
    Text Prompt Ensemble & 69.67 & 50.17 & 67.25 & 52.48 \\
    Exhaustive per-image VQA & 66.92 & 63.84 & 67.68 & 66.97 \\
    \textbf{ProbeScout initial $F_0$} & \textbf{86.95} & \textbf{72.21} & \textbf{86.14} & \textbf{72.70} \\
    \bottomrule
  \end{tabular}
\end{table}

To complement the VQA-budget curves, we compare sparse supervision followed by probe-based retrieval with exhaustive per-image VQA and the image- and text-embedding baselines (RQ1, RQ2). Annotation cost limits this additional comparison to 10 tasks: all eight Stanford Cars tasks in the candidate pool and \emph{Bike jumping} and \emph{Bicycle riding (no jump)} from HICO-DET. This set includes \emph{Dodge wagon}, whose reference annotations were corrected following full-gallery VQA; it is not part of the 17-task main evaluation. Exhaustively labeling CelebA, with approximately 200,000 images, would substantially increase the VQA cost. The selected tasks provide a direct comparison across fine-grained vehicle and human--object interaction retrieval. All four methods are evaluated on these same tasks, with equal task weights. ProbeScout uses its initial $F_0$ configuration before human feedback, with VQA supervision on at most 2\% of each task's gallery images.

On these tasks, ProbeScout achieves the highest AP and F1 on both Test and Gallery (\cref{tab:exhaustive-vqa}). Its Test AP exceeds Query MaxSim and exhaustive VQA by 9.14 and 20.03 percentage points, respectively. Compared with exhaustive VQA, its Test and Gallery F1 improve by 8.37 and 5.73 percentage points. One possible explanation for the AP gap is that discrete VQA confidence scores provide limited granularity for cross-image ranking. The F1 gains also support improved relevance discrimination. These results show that sparse VQA supervision combined with attribute probes can reduce image-labeling requirements while improving ranking and screening on the evaluated tasks.

\subsubsection{Core Component Ablations}
\label{sec:component-ablations}
\begin{table}[tb]
  \caption{Core component ablations on the 17 evaluation tasks. Values are task-macro AP and F1 (\%). Bold indicates the best value in each column.}
  \label{tab:component-ablations}
  \centering\normalsize
  \setlength{\tabcolsep}{3pt}
  \begin{tabular}{@{}lrrrr@{}}
    \toprule
    & \multicolumn{2}{c}{Test} & \multicolumn{2}{c}{Gallery} \\
    \cmidrule(lr){2-3}\cmidrule(l){4-5}
    Configuration & AP & F1 & AP & F1 \\
    \midrule
    \textbf{Full ProbeScout} & \textbf{83.21} & \textbf{72.27} & 82.33 & \textbf{73.65} \\
    w/o SoftGate & 82.63 & 71.29 & \textbf{82.47} & 72.11 \\
    w/o embedding branch & 81.58 & 71.22 & 80.57 & 72.95 \\
    w/o both & 81.73 & 70.93 & 81.51 & 71.97 \\
    \bottomrule
  \end{tabular}
\end{table}

To address RQ1, \cref{tab:component-ablations} compares full ProbeScout with configurations that remove SoftGate, the query-level embedding branch, or both, on the 17 evaluation tasks.
Full ProbeScout achieves the highest Test AP (83.21\%), Test F1 (72.27\%), and Gallery F1 (73.65\%). Removing either component lowers both Test metrics, supporting the complementary roles of attribute gating and query-level evidence in ranking quality and relevance discrimination. The configuration without SoftGate has slightly higher Gallery AP (82.47\% versus 82.33\%).

\subsubsection{Human Feedback and Sample Selection}
\label{sec:feedback-selection-evaluation}
\begin{table}[tb]
  \caption{Task-macro Test AP (\%) of feedback strategies on 17 tasks. Human-operated conditions are averaged across operators within each task, then equally across tasks, including the untuned Ford Mustang task. All refinement uses staged weight updates with frozen probes. Complete results are in Appendix~\ref{appendix:feedback-results}.}
  \label{tab:feedback-ap}
  \centering\normalsize
  \setlength{\tabcolsep}{3pt}
  \begin{tabular}{@{}lr@{}}
    \toprule
    Strategy & Test AP \\
    \midrule
    Initial $F_0$ & 83.21 \\
    Direct Tune & 82.18 \\
    Top-$K$ only & 82.80 \\
    Diagnostic + VQA & 83.42 \\
    Full ProbeScout & \textbf{85.18} \\
    \bottomrule
  \end{tabular}
\end{table}

\paragraph{Feedback strategies and human-guided analysis (RQ3, RQ4).}
\Cref{tab:feedback-ap} summarizes macro Test AP, while Appendix~\ref{appendix:feedback-results} reports all per-task results. We compare the initial ranking with Direct Tune (no additional human feedback), Top-$K$ only, Diagnostic + VQA (four diagnostic filters with automatic VQA labels), and full ProbeScout on the 17 evaluation tasks. The full workflow combines PCP evidence, rank-profile clusters, visual-embedding clustering, and diagnostic candidates with human judgments. All refinement uses staged weight updates with frozen probes.

Interactive runs were limited to 10 minutes of analysis per task. Each task was performed by three to five researchers from different fields, including AI and non-AI backgrounds. For human-operated conditions, AP was averaged across operators within each task, followed by an equally weighted macro average across the 17 tasks. \emph{Ford Mustang convertible two-door} has few, readily retrieved positives and already achieves 100\% initial AP. This task was not tuned; its initial result is retained across strategy columns and included in the macro average.

The full workflow raises macro Test AP from 83.21\% to 85.18\%, exceeding Direct Tune, Top-$K$ only, and Diagnostic + VQA by 3.01, 2.38, and 1.77 percentage points, respectively; differences are calculated before rounding. It achieves the best or tied-best result in 14 of the 17 tasks across all five configurations, and matches or exceeds both Top-$K$ only and Diagnostic + VQA on every task. The largest gain over the initial ranking is 12.34 percentage points for \emph{BMW sedan}. Test AP decreases for \emph{Directing airplane} (84.15\% to 79.96\%) and \emph{Herding cow} (65.11\% to 64.42\%).
These results indicate that coordinated analysis adds value to an already effective initial retrieval model. Gains can be substantial when the initial ranking is weaker, as in \emph{BMW sedan}, while strong starting performance can also improve: \emph{Hyundai hatchback} rises from 89.68\% to 93.08\% (+3.40\%), and \emph{Shearing sheep} gains 1.14 points from an initial AP above 90\%. The higher macro AP than Direct Tune, Top-$K$ only, and Diagnostic + VQA supports combining visual evidence with human judgments beyond weight optimization, head-of-ranking inspection, or automated diagnostic labeling alone. The case studies (\cref{sec:case-studies}) illustrate how analysts use these capabilities to identify informative candidates and confirm unmet attributes.

\subsection{Case Studies}
\label{sec:case-studies}
\begin{figure}[tbp]
  \centering
  \includegraphics[width=\columnwidth,alt={Two recorded case studies. Sheep-shearing retrieval shows failed-attribute confirmation, rollback, intersection-based error discovery, and AP outcomes in panels a through d. Hyundai hatchback retrieval shows failed-attribute correction, sedan and SUV discovery, and AP outcomes in panels e through h.}]{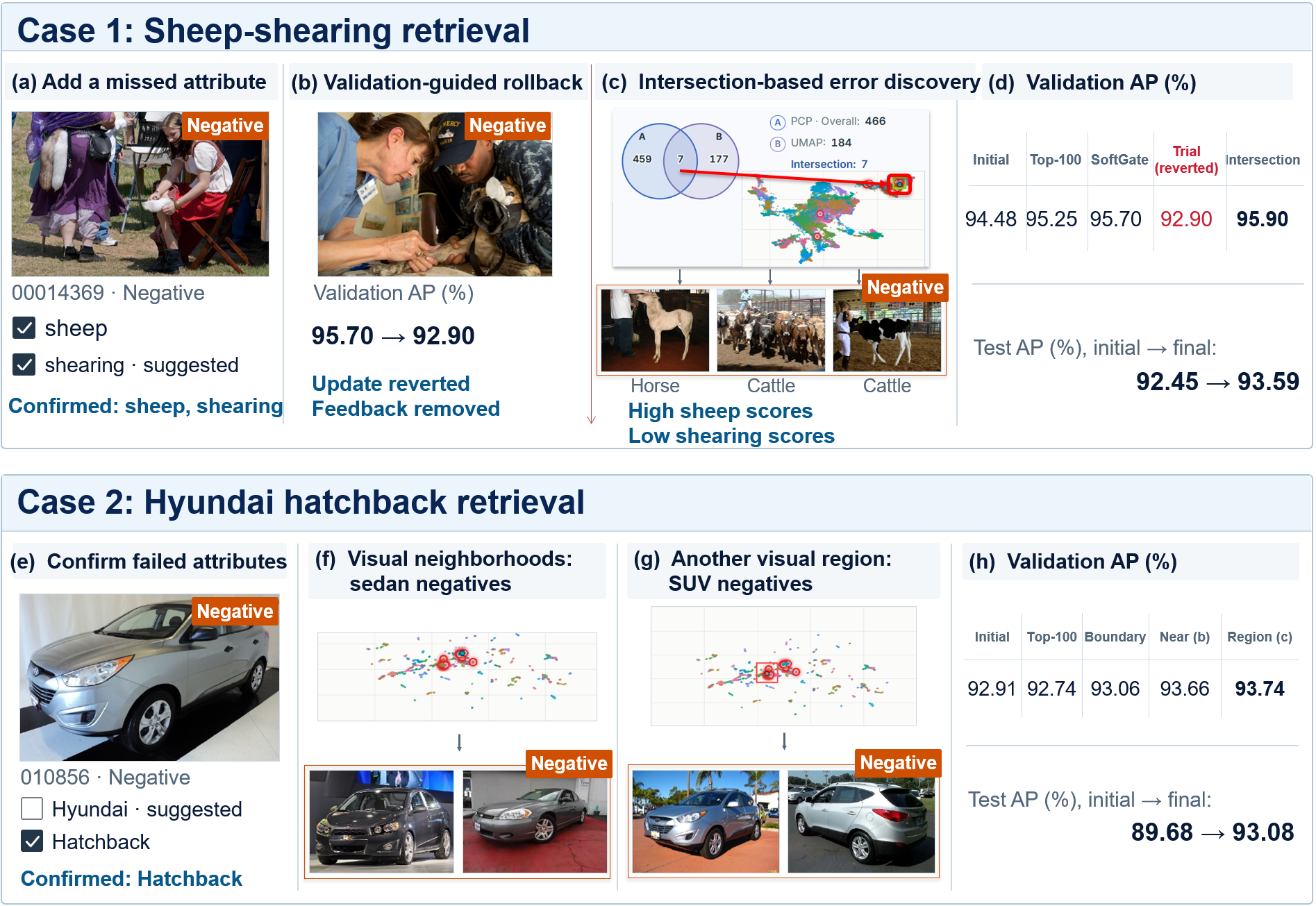}
  \caption{Selected interactions and outcomes in two case studies. Sheep-shearing retrieval: adding a missed failure attribute (a), validation-guided rollback (b), intersection-based error discovery (c), and Val/Test AP (d). Hyundai hatchback retrieval: failed-attribute correction (e), sedan-neighborhood inspection (f), SUV discovery (g), and Val/Test AP (h). Red circles mark labeled images; outlined regions denote selections.}
  \label{fig:case-studies}
\end{figure}

We present two illustrative cases selected from repeated trials with researchers from AI and non-AI fields. The sessions were chosen for the clarity with which they show complementary system functions: inspecting attribute evidence and correcting supervision in sheep-shearing retrieval, and expanding feedback from local errors to related image groups in Hyundai hatchback retrieval. The first operator had an AI background and retrieval experience, while the second had experience using related systems. 
All updates used staged weight refinement with frozen probes. \cref{fig:case-studies} presents key interactions and outcomes.

\subsubsection{Sheep-Shearing Retrieval}
\label{sec:case-shearing}
A participant with a background in AI and experience in image retrieval used the attributes \emph{sheep} and \emph{shearing} to retrieve sheep-shearing images from HICO-DET. The session lasted approximately 17 minutes: two minutes for reading the task, two minutes for familiarization with the system, and 13 minutes for interactive analysis and refinement. This illustrative session was not included in the time-bounded comparison in \cref{sec:feedback-selection-evaluation}. System responses took only seconds and accounted for a negligible portion of the session. The participant first inspected the top 100 results and added two positive and five negative images as feedback. Refinement increased Val AP from 94.48\% to 95.25\%. Adding two further feedback images through the SoftGate boundary filter and refining again raised Val AP to 95.70\%. During inspection, the participant identified a high-scoring false positive containing no sheep (\cref{fig:case-studies}(a)). Although the system suggested only \emph{shearing} as the failed attribute, the participant confirmed that both \emph{sheep} and \emph{shearing} were unmet, extending negative supervision to both conditions.

A subsequent update incorporating one additional feedback image reduced Val AP from 95.70\% to 92.90\%, a decrease of 2.80 percentage points (\cref{fig:case-studies}(b)). The participant reverted the update, removed the image, and continued inspecting the remaining errors. The participant combined three candidate sets: images brushed in the visual embedding view, SoftGate-filtered candidates, and images selected by their attribute-score profiles in the PCP. After examining set overlaps in the Venn view, the participant inspected the intersection candidates with high \emph{sheep} scores but low \emph{shearing} scores in the linked gallery (\cref{fig:case-studies}(c)). These included horse and cattle images that received high \emph{sheep} scores despite failing the object condition. Linked inspection thus extended an individual false positive into a group of object-level errors sharing a similar score pattern.

Further inspection revealed an image labeled positive for \emph{sheep} but negative for \emph{shearing} by VQA. The person was holding a sheep and had scissors in hand, with wool visible on the ground. Based on these cues, the participant judged that the image satisfied the \emph{shearing} condition, retained the positive \emph{sheep} label, and changed \emph{shearing} from negative to positive before including the image in the supervision set. This phase of exploration added ten feedback images in total, and the participant retained the configuration obtained after refinement.

Overall, Val AP increased from 94.48\% to 95.90\%, while Test AP increased from 92.45\% to 93.59\%, a gain of 1.14 percentage points (\cref{fig:case-studies}(d)). The case illustrates how linked inspection connected attribute scores with image content to identify object-level false positives and correct a VQA-derived action label. Validation and rollback supported discarding an update that reduced performance and continuing to revise the feedback set.

\subsubsection{Hyundai Hatchback Retrieval}
\label{sec:case-hyundai}

Another participant with experience using related systems received an introduction to ProbeScout's functionality and then used example images and the attributes \emph{Hyundai} and \emph{Hatchback} to retrieve Hyundai hatchbacks from Stanford Cars. The session lasted approximately 10 minutes, with three minutes spent reading the task and becoming familiar with the system, and the remaining seven minutes devoted to the retrieval task. The participant first inspected the top 100 results and added eight feedback images, marking body-type mismatches as negative examples. The first refinement slightly reduced Val AP from 92.91\% to 92.74\%. Negative feedback also involved confirming the failed condition. For example, the Hyundai SUV in \cref{fig:case-studies}(e) satisfied the brand requirement but not the hatchback requirement. The participant changed the failed attribute from the system-suggested \emph{Hyundai} to \emph{Hatchback}, directing negative supervision to the unmet body-type condition.

The visual embedding view showed that the labeled images were concentrated in one region. The participant then inspected SoftGate boundary candidates, identified additional sedans, and added eight more feedback images. Refinement raised Val AP to 93.06\%. Next, the participant highlighted labeled images and inspected their visual neighborhoods through the linked gallery, finding further sedans with similar appearances (\cref{fig:case-studies}(f)). This neighborhood inspection contributed nine additional feedback images, and subsequent refinement increased Val AP to 93.66\%.
The participant then brushed another dense region and found two SUVs among the first ten displayed candidates that had not yet been selected for feedback (\cref{fig:case-studies}(g)). Labeling them as negative and refining again raised Val AP to 93.74\%. The four rounds leading to this configuration added 27 feedback images. Subsequent trials did not further improve Val AP, so the participant retained this configuration for final evaluation.

Overall, Val AP increased from 92.91\% to 93.74\%, while Test AP increased from 89.68\% to 93.08\%, a gain of 3.40 percentage points (\cref{fig:case-studies}(h)). The case illustrates how diagnostic filtering and linked embedding--gallery inspection extended feedback collection from individual body-type mismatches to related sedan images and SUVs in another visual region. Failed-attribute confirmation translated negative examples that satisfied the brand condition into attribute-level supervision for the body-type requirement.

\section{Discussion}
\label{sec:discussion}

\paragraph{Human--AI Division of Labor}
ProbeScout assigns collection-wide scoring and organization to the machine, while analysts define conditions, interpret evidence, and decide which images belong in a useful subset.
Failed-attribute suggestions require human confirmation because a weak model response does not necessarily identify the unmet visual condition, as illustrated in \cref{sec:case-hyundai}. Model explanations can increase acceptance of recommendations without improving their correctness~\cite{bansal2021complementary}. Accordingly, failed-attribute suggestions remain editable and uncertain judgments need not become supervision, consistent with user-correction guidelines~\cite{amershi2019guidelines}.

\paragraph{Transferable Design Implications}
The design suggests three principles for systems that compose reusable evidence. First, \emph{make complementary grouping criteria inspectable}: rank-profile and visual-embedding patterns answer different questions, and concrete examples help analysts interpret the difference. Second, \emph{connect diagnosis to revisable selections}: brushing, filters, and overlap summaries should lead to bounded image sets whose origins remain visible. Third, \emph{separate exploration from supervision}: analysts retain control over which inspected images become a target subset and which confirmed judgments enter refinement.

\paragraph{Generalizability and Scalability}
The workflow applies when collection-analysis goals decompose into conditions with inspectable per-image evidence. Its evidence representation accommodates different frozen encoders. Subtle relations, spatial configurations, and domain-dependent concepts require particular care in attribute definition and verification; consistent semantics are essential when reusing an attribute across tasks.
Reuse is strongest when evolving subset goals recombine familiar conditions on compatible features and galleries. A new visual attribute extends the evidence through offline VQA acquisition and probe training; interactive refinement keeps the resulting probes fixed. Changes to encoders, domains, or attribute definitions may require refreshing scores or calibration. At larger scales, storage, clustering, projection, and display contribute to interaction cost alongside score computation.

\paragraph{Limitations and Future Work}
Correlated encoder and VQA errors can affect multiple probes, while PU ordering may penalize unlabeled positives and active selection may bias calibration. Projection distortion~\cite{jeon2025backstage} and incomplete inspection can also leave relevant groups undiscovered. These issues motivate coverage-aware inspection and closer examination of uncertain evidence.
Staged weight refinement preserves the soft-conjunction structure, with its corrective capacity bounded by the distinctions represented in the frozen evidence. Linked image inspection supports verification of the resulting subsets. Repeated use of the single VQA Validation set can also lead to adaptive overfitting, a concern studied in repeated holdout evaluation~\cite{blum2015ladder}. Future work will examine context-sensitive attribute definitions, criteria for extending the evidence, and the tradeoffs among subset quality, analytical effort, and feedback in the planned workflow studies.

\section{Conclusion}
\label{sec:conclusion}
We presented ProbeScout, a visual analytics system for attribute-guided screening and subset construction in image collections. The system combines reusable multi-probe evidence with analyst-directed selection and staged weight refinement. Benchmark comparisons and VQA-budget experiments on the same 17 tasks demonstrate improved retrieval quality and efficient acquisition. Human-guided analysis further improves macro Test AP from 83.21\% to 85.18\%. A separate analysis of new query combinations validates attribute-probe reuse without retraining. Two illustrative case studies show how linked evidence inspection and confirmed judgments support targeted refinement. 

\FloatBarrier
\bibliographystyle{abbrv-doi-hyperref}
\bibliography{references}

\clearpage
\appendix
\crefalias{section}{appendix}
\twocolumn[{%
\begin{minipage}{\textwidth}
\section{Retrieval Tasks and Retained Query Attributes}
\label{appendix:task-scope}
\paragraph{Task sampling and annotation review.}
Candidate tasks were initially sampled at random from six source datasets, subject to the presence of positive images satisfying the required attribute conjunction. Retrieval experiments and linked image inspection subsequently exposed errors in some original GT attribute annotations. Following annotation review, some candidate tasks were excluded because of problems in their evaluation labels, with image-level evidence recorded for each exclusion. The retained main evaluation contains 17 tasks from three datasets, listed below. Appendix~\ref{appendix:annotation-quality} illustrates potential annotation discrepancies surfaced by the system.

\begingroup
  \captionof{table}{Retrieval tasks and retained query attributes. The main evaluation comprises 17 tasks across three datasets. Semicolons separate distinct modeled attributes.}
  \label{tab:task-attributes}
  \centering\normalsize
  \setlength{\tabcolsep}{6pt}
  \renewcommand{\arraystretch}{1.08}
  \begin{tabular}{@{}p{.36\textwidth}p{\dimexpr.64\textwidth-12pt\relax}@{}}
    \toprule
    Task & Retained query attributes \\
    \midrule
    \multicolumn{2}{@{}l}{\textbf{Stanford Cars (7 tasks)}} \\
    BMW convertible & BMW brand identity; Convertible body type with soft-top roof \\
    BMW sedan & BMW kidney grille; sedan body shape \\
    Ford Mustang convertible two-door & Ford brand identity; Mustang model family; Convertible body type \\
    Hyundai hatchback & Hyundai; Hatchback \\
    Honda minivan & Honda; Minivan \\
    Jeep SUV & Jeep; SUV \\
    GMC van & GMC; Van \\
    \midrule
    \multicolumn{2}{@{}l}{\textbf{HICO-DET (8 tasks)}} \\
    Bike jumping & bicycle; jumping \\
    Bicycle riding (no jump) & bicycle; riding; not jumping; not bike jump \\
    Shearing sheep & sheep; shearing \\
    Typing on keyboard & keyboard; typing \\
    Directing airplane & airplane; directing \\
    Sailing boat & boat; sailing \\
    Herding cow & cow; herding \\
    Reading a book & book; reading \\
    \midrule
    \multicolumn{2}{@{}l}{\textbf{CelebA (2 tasks)}} \\
    Gray hair + eyeglasses + male & Gray\_Hair; Eyeglasses; Male \\
    Eyeglasses, smiling, and necktie & Eyeglasses; Smiling; Wearing\_Necktie \\
    \bottomrule
  \end{tabular}

  \medskip
  \parbox{\textwidth}{The BMW rows list only the attributes used for retrieval. The Ford task name includes ``two-door,'' but this is not modeled as a separate attribute. For bicycle riding, ``not jumping'' and ``not bike jump'' are distinct modeled attributes.}
\endgroup

\end{minipage}
}]

\twocolumn[{%
\begin{minipage}{\textwidth}
\section{Inspecting Ground-Truth Attribute Annotations}
\label{appendix:annotation-quality}
Beyond retrieval refinement, ProbeScout can also help identify potential errors in ground-truth (GT) attribute annotations. By linking retrieval results with attribute evidence, analysts can locate images whose visible content conflicts with their original labels. For example, \cref{fig:gt-annotation-audit}(c) shows a person wearing eyeglasses despite a negative eyeglasses label. Such discrepancies can guide manual review of potentially incorrect or missing annotations and support data-quality improvement.

\medskip
\begin{center}
  \includegraphics[width=.94\textwidth,height=.75\textheight,keepaspectratio,alt={Four examples of potential ground-truth annotation discrepancies, with retrieval views on the left and enlarged images and original labels on the right: typing on a keyboard, hugging a cat, gray hair with eyeglasses and male, and bald with smiling and necktie.}]{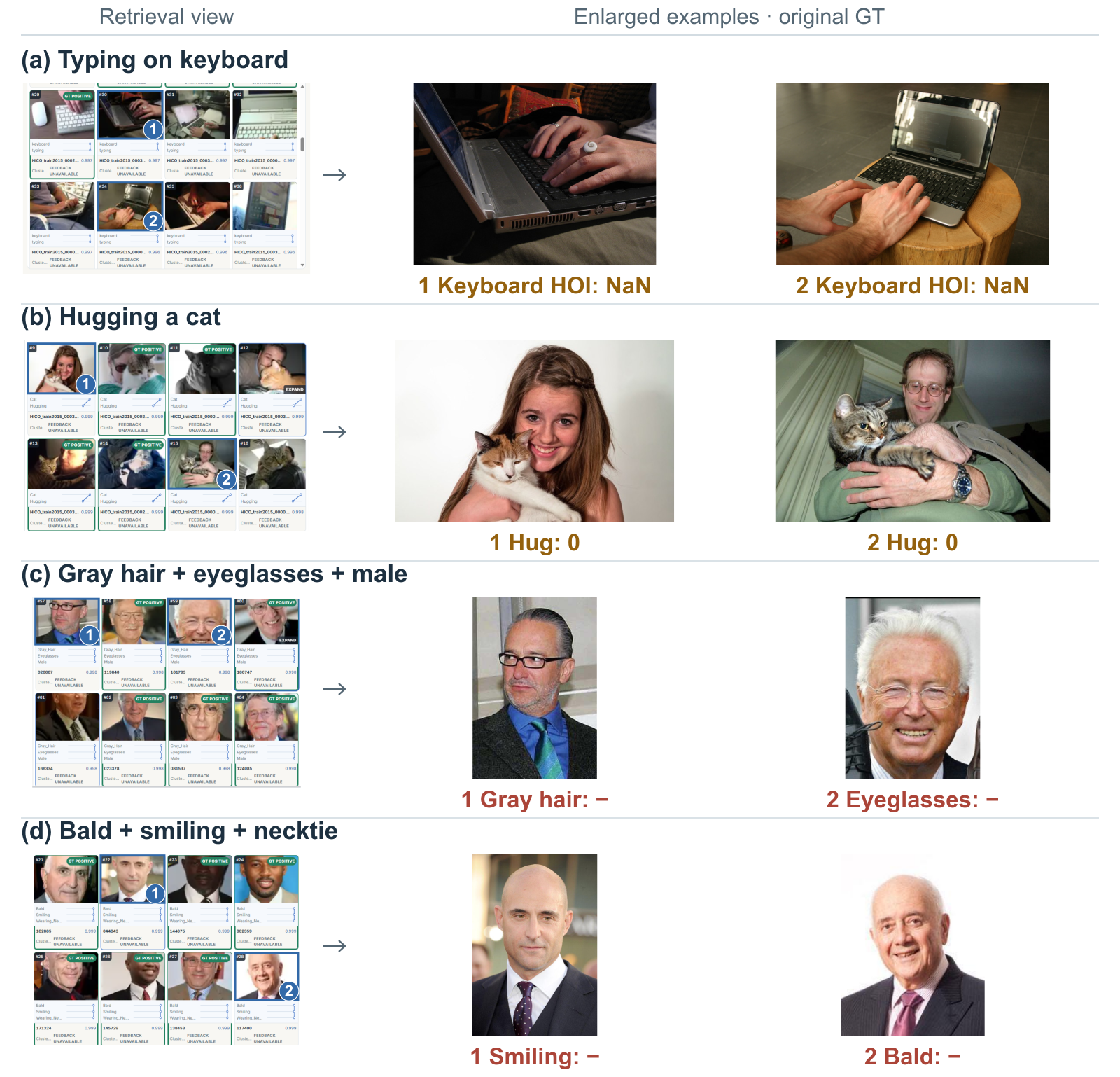}
  \captionof{figure}{Examples of potential discrepancies in original GT annotations surfaced through ProbeScout. Each panel shows retrieval results (left) and enlarged examples with their original labels (right): (a) typing on a keyboard; (b) hugging a cat; (c) gray hair, eyeglasses, and male; and (d) bald, smiling, and necktie. HOI denotes human--object interaction.}
  \label{fig:gt-annotation-audit}
\end{center}
\end{minipage}
}]

\clearpage
\section{Implementation and Optimization Details}
\label{appendix:implementation}
This appendix records detailed configurations of 8 probes in \cref{sec:sparse-verification}, implementation settings for the protocol in \cref{sec:evaluation}, and the staged refinement in \cref{sec:refinement}.

\subsection{Attribute Probe Configurations}
\label{appendix:probe-configurations}

We use eight lightweight probe configurations on frozen visual
representations. Six operate on pooled image features, while the
two attention-based configurations use patch-token features.

\paragraph{Supervised configurations.}
MLP uses class-weighted supervision.
K-Fold reweights positive labels using out-of-fold confidence.
Triplet Loss augments supervised learning with a triplet-margin
objective~\cite{schroff2015facenet}.

\paragraph{Attention-based configurations.}
Attention Pooling uses learned-query pooling over patch features.
Attribute-conditioned Attention conditions patch pooling on the
attribute's text representation.
Both use class-weighted supervision.

\paragraph{PU-inspired classification configurations.}
Positive--unlabeled learning estimates classification risk from
positive and unlabeled examples~\cite{duplessis2015convexpu}.
Our nnPU-inspired configuration~\cite{kiryo2017nnpu} adds a
nonnegative unlabeled-risk correction to labeled positive and
negative losses.
The DC-PU-inspired variant~\cite{li2025dcpu} additionally penalizes
imbalance between positive and corrected negative risks.
Both incorporate available labeled negatives and are therefore
adaptations of the cited methods; they are denoted nnPU and DC-PU
in the interface and experiments.

\paragraph{PU-ranking configuration.}
PURA combines labeled supervision with a soft pairwise ranking
loss over positive--unlabeled pairs.

\subsection{Models and Probe Training}
We use a frozen SigLIP ViT-B/16 visual encoder~\cite{zhai2023siglip} with $224\times224$ inputs, producing 768-dimensional image features; attention-based probes use $196\times768$ patch-token features. Qwen-VL-Max (API model ID \texttt{qwen-vl-max})~\cite{alibaba2026qwenvlapi} is the default model for query-attribute extraction and attribute labeling of candidate images. Each attribute has eight probes, trained for 100 epochs with batch size 64, initial learning rate $10^{-3}$, weight decay $10^{-4}$, and cosine learning-rate scheduling. We use five random seeds (0--4). Each run completes all 100 epochs, and the checkpoint with the highest area under the receiver operating characteristic curve (ROC-AUC) on fixed Val is selected; training is not stopped early.

\paragraph{Initial fusion.}
The initial configuration $F_0$ uses $\beta_{am}=1/8$, $\gamma_a=1$, $\eta_1=\eta_2=1/2$, and $\lambda_0=0.25$, with gate parameters selected as specified below.

\subsection{Clustering and Projection}
\label{sec:clustering-settings}
Fine-grained grouping uses mini-batch $k$-means~\cite{sculley2010webscale} with 30, 50, or 100 clusters (default: 50; random seed: 42), applied to per-axis standardized rank-profile features or L2-normalized 768-dimensional visual embeddings. Preprocessing, clustering, and projection models are fitted on Development images; cluster membership is computed in these feature spaces rather than in the displayed two-dimensional projection. For display, UMAP uses cosine distance after PCA reduction to at most 50 dimensions, with 30 neighbors and a minimum distance of 0.1.

\subsection{Budget Schedule, Calibration, and Statistics}
\label{appendix:calibration-settings}
The VQA-budget comparison uses 50, 100, 150, 200, and 300 images per task. Iterative acquisition starts with 100 images and adds 40 fitting images and 10 audit images per round. The 50-image condition uses the stratified initialization subset described in \cref{sec:sparse-verification}. Adaptive stopping is disabled for these fixed-budget comparisons.

For each configuration with SoftGate, candidate gates are constructed from the stored attribute-calibration values $\theta_a^{\mathrm{ref}}$ and $T_a^{\mathrm{ref}}$. A candidate uses $\theta_a=\operatorname{clip}(\theta_a^{\mathrm{ref}}+\delta,0,1)$ and $T_a=\min(\kappa T_a^{\mathrm{ref}},0.30)$, where $\delta\in\{0,\pm0.05,\pm0.10\}$ and $\kappa\in\{1,1.5,2,3\}$ are shared across attributes within each task. Val AP selects the candidate, with fusion weights held at their initial values. The Fusion and Full ProbeScout rows in \cref{tab:probe-performance,tab:component-ablations} use identical parameters and scores. In the separate VQA-budget experiment, AP is averaged over five seeds within each task and then equally across tasks; 95\% confidence intervals use 10,000 paired task-bootstrap resamples~\cite{efron1979bootstrap}.

\subsection{Staged Refinement Specification}
\label{appendix:optimization}
\paragraph{Feedback supervision.}
For image $i$, feedback is encoded for training as a signed weight $\ell_i\in\{-16,-8,0,+8,+16\}$, denoting strong negative, negative, uncertain, positive, and strong positive, respectively. The sign determines the binary relevance label, and the magnitude specifies supervision strength. Zero preserves any existing VQA label and adds no feedback supervision. A positive query judgment confirms all required attributes. For each negative judgment used in refinement, the analyst confirms at least one failed attribute, aided by a suggestion based on the weakest initial gate. Confirmed failures receive negative attribute labels; other attributes retain existing VQA labels or remain unknown (R4).

Refinement combines accumulated eligible DG feedback for the current query with the original VQA fitting supervision. Human judgments override the corresponding original labels on repeated images. Original VQA rows have raw weight one; human feedback rows use $|\ell_i|$, giving weights 8 and 16 for ordinary and strong feedback. These weights are rescaled to equalize positive and negative total mass separately for each attribute and for the query. Attribute judgments retain their named conditions; complete-query evidence and its fusion parameters retain the current query scope.

\paragraph{Two-stage optimization.}
Both stages use the score model in \cref{eq:attribute-gate,eq:initial-ranking}, with the backbone, all learned probes, normalization statistics, and SoftGate thresholds and temperatures frozen. Each run starts from the same initial fusion configuration and uses the accumulated supervision, rather than warm-starting from previously fitted weights. Let $\Omega_A$ be the set of image--attribute pairs with known training labels $\widetilde y_{ia}$, and $\Omega_Q$ the image indices with known query labels $\widetilde y_i^q$ after supervision merging. Denote their balanced weights by $\omega_{ia}$ and $\omega_i$, respectively. The class-balanced weighted BCE losses for attribute and query supervision are
\begin{equation}
\begin{aligned}
\mathcal L_A&=\frac{\sum_{(i,a)\in\Omega_A}\omega_{ia}
\operatorname{BCE}(\widetilde y_{ia},g_{ia})}
{\sum_{(i,a)\in\Omega_A}\omega_{ia}},\\
\mathcal L_Q&=\frac{\sum_{i\in\Omega_Q}\omega_i
\operatorname{BCE}(\widetilde y_i^q,F_i)}
{\sum_{i\in\Omega_Q}\omega_i}.
\end{aligned}
\label{eq:refinement-losses}
\end{equation}

Let $K_q=|\mathcal A_q|$ be the number of active attributes, let $\rho_A,\rho_Q\geq0$ weight the two data losses, and let $\zeta_\beta,\zeta_\gamma,\zeta_\eta,\zeta_\lambda\geq0$ be regularization coefficients. Stage 1 optimizes only the probe-level weights:
\begin{equation}
\min_{\{\beta_a\}}\quad \rho_A\mathcal L_A+
\frac{\zeta_\beta}{K_q}\sum_{a\in\mathcal A_q}\sum_{m=1}^{M}
\left(\beta_{am}-\frac{1}{M}\right)^2.
\label{eq:stage-one}
\end{equation}
The remaining fusion parameters are fixed during this stage.

Stage 2 fixes the probe weights and optimizes the attribute exponents and embedding branch, with the negative-example gradient routing specified below:
\begin{equation}
\begin{aligned}
\min_{\gamma,\eta,\lambda}\quad
&\rho_Q\mathcal L_Q+\zeta_\gamma\sum_{a\in\mathcal A_q}(\gamma_a-1)^2\\
&+\zeta_\eta\sum_{k=1}^{2}(\eta_k-1/2)^2
+\zeta_\lambda(\lambda-\lambda_0)^2.
\end{aligned}
\label{eq:stage-two}
\end{equation}
Here $k$ indexes the two query-level embedding scorers. Softmax parameterizations preserve the simplex constraints on each $\beta_a$ and on $\eta$, and a sigmoid keeps $\lambda$ strictly between zero and one during fitting. Each $\gamma_a$ is independently sigmoid-bounded by fixed limits $0<\gamma_{\min}<1<\gamma_{\max}$. Positive query examples supply query-loss gradients to all attribute exponents; negative examples supply them only to exponents of failed attributes identified by VQA labels or confirmed in human feedback. This per-example routing preserves forward scores and loss values, with regularization unchanged. Both positive and negative examples supply gradients to the embedding branch.

The system retains the candidate with the lowest combined regularized training objective among the initial state, completed Stage-1 state, and completed Stage-2 state, then reports its ranking and fixed-Val metrics for analyst comparison.

\paragraph{Weight-refinement settings.}
Staged weight refinement uses Adam with learning rate 0.05 for 500 steps in total, split equally between the two stages (250 steps each). Before class balancing, original VQA labels have raw weight 1, ordinary human feedback has weight 8, and strong feedback has weight 16, as defined by $|\ell_i|$ above. The loss coefficients are $\rho_A=\rho_Q=1$. All four anchor regularization coefficients are 0.05: $\zeta_\beta=\zeta_\gamma=\zeta_\eta=\zeta_\lambda=0.05$. Attribute exponents satisfy $0.05<\gamma_a<3$.

\twocolumn[{%
\begin{minipage}{\textwidth}
\section{Complete Per-Task Results}
\label{appendix:complete-results}
\subsection{Attribute Reuse}
\label{appendix:reuse-results}
\Cref{tab:probe-reuse-full} expands the macro comparison in \cref{tab:probe-reuse} into all 12 new query combinations. Existing attribute probes are reused without retraining or new query-level labels.
\begin{center}
  \captionof{table}{Attribute reuse on 12 new query combinations. Test AP and F1 are percentages. F1 uses original attribute-training thresholds followed by AND; the macro average weights tasks equally.}
  \label{tab:probe-reuse-full}
  \centering\normalsize
  \setlength{\tabcolsep}{3pt}
  \begin{tabular}{@{}lrrrr@{}}
    \toprule
    & \multicolumn{2}{c}{\shortstack{ProbeScout\\Reuse}}
    & \multicolumn{2}{c}{\shortstack{Attribute-text\\mean}} \\
    \cmidrule(lr){2-3}\cmidrule(l){4-5}
    Task & AP & F1 & AP & F1 \\
    \midrule
    Dodge SUV & 86.22 & 24.30 & 49.25 & 13.79 \\
    Ford Sedan & 77.14 & 93.33 & 41.60 & 15.22 \\
    Hyundai Sedan & 99.58 & 90.91 & 67.81 & 61.11 \\
    Hyundai SUV & 91.74 & 84.21 & 71.14 & 40.00 \\
    Hug dog & 74.24 & 50.00 & 65.78 & 31.25 \\
    Jump horse & 73.02 & 66.67 & 58.09 & 19.67 \\
    Jump motorcycle & 78.62 & 53.85 & 36.77 & 42.11 \\
    Jump skateboard & 64.20 & 66.67 & 79.20 & 68.38 \\
    Bald + Male & 66.19 & 67.93 & 28.46 & 45.38 \\
    Bangs + Smiling & 64.68 & 66.44 & 26.05 & 38.42 \\
    Black Hair + Male & 60.76 & 47.65 & 42.62 & 48.12 \\
    Eyeglasses + Wearing Hat & 70.77 & 52.27 & 21.22 & 41.34 \\
    \midrule
    \textbf{Macro average} & \textbf{75.60} & \textbf{63.69} & \textbf{49.00} & \textbf{38.73} \\
    \bottomrule
  \end{tabular}
\end{center}

\subsection{Feedback Strategies}
\label{appendix:feedback-results}
\Cref{tab:feedback-ap-full} expands the macro comparison in \cref{tab:feedback-ap} into all 17 tasks, using the 10-minute analysis protocol in \cref{sec:feedback-selection-evaluation}. All refinement uses staged updates with frozen probes; the untuned Ford Mustang task remains included in the macro average.
\begin{center}
  \captionof{table}{Test AP (\%) of feedback strategies on 17 tasks. Direct Tune adds no human feedback; Diagnostic + VQA uses four diagnostic filters with automatic VQA labels; full ProbeScout combines visual analysis and human judgments. Human-operated results are averaged across operators within each task. The Ford Mustang task was not tuned: all columns retain its initial result. Tasks are equally weighted in the macro average. Bold marks row-wise best results, including ties.}
  \label{tab:feedback-ap-full}
  \centering\normalsize
  \setlength{\tabcolsep}{3pt}
  \begin{tabular}{@{}lrrrrr@{}}
    \toprule
    Task & Initial $F_0$ & Direct Tune & Top-$K$ only
         & \shortstack[r]{Diagnostic\\+ VQA} & \shortstack[r]{Full\\ProbeScout} \\
    \midrule
    BMW convertible & 99.88 & 99.63 & 99.64 & 99.73 & \textbf{99.91} \\
    BMW sedan & 80.50 & 81.57 & 83.70 & 85.31 & \textbf{92.84} \\
    Ford Mustang convertible two-door & \textbf{100.00} & \textbf{100.00} & \textbf{100.00} & \textbf{100.00} & \textbf{100.00} \\
    Bike jumping & 83.12 & 85.19 & 85.46 & 85.29 & \textbf{86.76} \\
    Bicycle riding (no jump) & 90.46 & 89.00 & 88.89 & \textbf{91.06} & \textbf{91.06} \\
    Shearing sheep & 92.45 & 92.49 & 91.89 & 91.79 & \textbf{93.59} \\
    Typing on keyboard & 72.99 & 68.95 & 74.20 & 73.16 & \textbf{74.95} \\
    Hyundai hatchback & 89.68 & 81.27 & 91.79 & 92.23 & \textbf{93.08} \\
    Directing airplane & \textbf{84.15} & 75.87 & 69.80 & 70.22 & 79.96 \\
    Sailing boat & 85.06 & \textbf{87.22} & 85.35 & 86.48 & 86.76 \\
    Herding cow & \textbf{65.11} & 60.98 & 61.05 & 62.83 & 64.42 \\
    Reading a book & 73.08 & 72.81 & 73.12 & 73.26 & \textbf{73.79} \\
    Honda minivan & 98.41 & 98.39 & 97.37 & 98.13 & \textbf{98.57} \\
    Jeep SUV & 99.14 & 98.38 & 98.43 & 98.30 & \textbf{99.41} \\
    GMC van & 83.33 & 81.40 & 83.20 & 85.61 & \textbf{86.36} \\
    Gray hair + eyeglasses + male & 64.39 & 69.63 & 69.68 & 68.26 & \textbf{69.73} \\
    Eyeglasses, smiling, and necktie & 52.77 & 54.22 & 54.08 & 56.43 & \textbf{56.93} \\
    \midrule
    \textbf{Macro average} & 83.21 & 82.18 & 82.80 & 83.42 & \textbf{85.18} \\
    \bottomrule
  \end{tabular}
\end{center}

\end{minipage}
}]

\twocolumn[{%
\begin{minipage}{\textwidth}
\section{Single-Attribute Retrieval through Probe Reuse}
\label{appendix:single-attribute-retrieval}
ProbeScout can reuse existing attribute probes to retrieve a single visual condition independently. The examples in \cref{fig:single-attribute-retrieval} use the initial model on CelebA Test. For each target, images are ranked by the corresponding attribute evidence without retraining or requiring the other attributes of the original composite query. These examples illustrate flexible changes of retrieval target through attribute-evidence reuse.

\medskip
\begin{center}
  \includegraphics[width=.94\textwidth,alt={Three rows of consecutive top-five CelebA Test results from the initial model: eyeglasses, smiling, and necktie. Columns show ranks one through five.}]{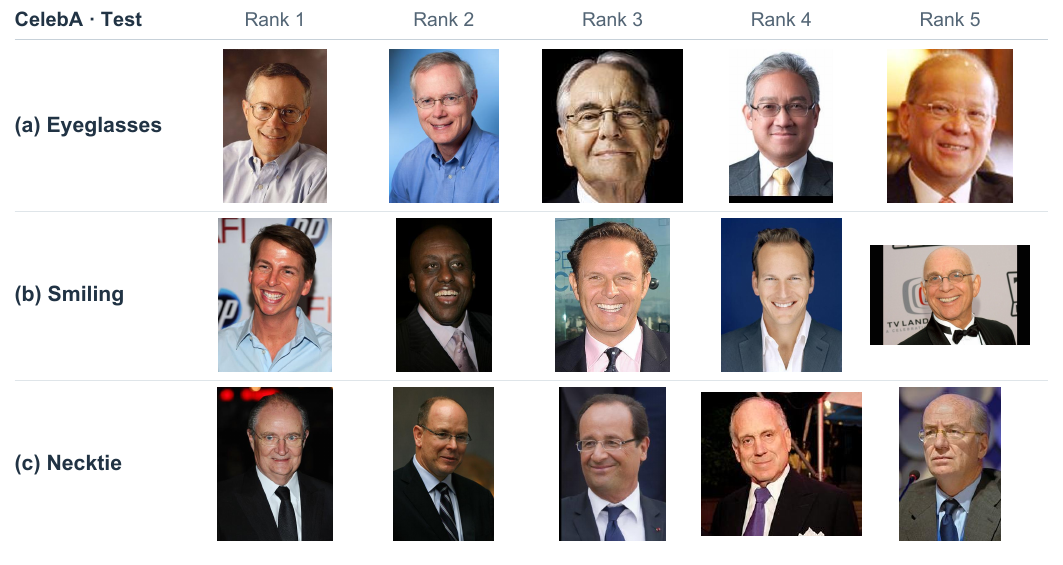}
  \captionof{figure}{Single-attribute retrieval on CelebA Test. Each row shows the consecutive top-five results from the initial model for (a) eyeglasses, (b) smiling, and (c) wearing a necktie, using the existing probes for that attribute.}
  \label{fig:single-attribute-retrieval}
\end{center}
\end{minipage}
}]

\end{document}